*Hypothesis*

# Electron Spin and the Origin of Bio-homochirality I. Extant Enzymatic Reaction Model

Wei Wang

CCMST, Academy of Fundamental and Interdisciplinary Sciences, Harbin Institute of Technology, Harbin, 150080, China

Email: wwang_ol@hite.edu.cn

Phone: 86-451-86418430

Fax: 86-451-86418440

**Abstract**

In this paper, I tentatively put forward a new hypothesis that the emergence of a single chiral form of biomolecules in living organisms is specifically determined by the electron spin state during their enzyme-catalyzed synthesis processes. Specifically speaking, the electrons released from the coenzyme NAD(P)H of amino acid synthase are heterogeneous in spin states; however, when they pass through the chiral α-helix structure of the enzymes to the site of amino acid synthesis at the other end of the helix, their spin states are filtered and polarized, producing only "spin up" electrons; once the spin-polarized electrons participate in the reductive reaction between α-oxo acid and ammonia, only L-amino acids are formed according to the Pauli exclusion principle.

## Background

As is well known, most biomolecules occur in mirror images of each other, while the essential molecules for life are overwhelmingly asymmetrical. Nature has wrought only one enantiomer in biology. Merely L-amino acids are incorporated into proteins, and D-sugars are preferentially consolidated into RNA and DNA.

How does this homochirality form in our body? To answer this question, here I would tentatively present a new hypothesis that the emergence of a single chiral form of biomolecules in living organisms is specifically determined by the electron spin state during their enzyme-catalyzed synthesis processes. How the specific electron spin state evolves in enzyme-catalyzed reactions and how it works on the formation of the chiral center of biomolecules are discussed, by using the biosynthesis of α-amino acid as a paradigm.

## Electron spin property determines chirality

In living organisms, reversible amination of α-oxo acids exists in the biosynthesis of all proteinic amino acids. Glutamate dehydrogenase and transaminase participate in the regulation of the equilibria among amino acids and α-oxo acids (Berg JM , Tymoczko JL  et al. 2002). In various bacteria, a similar function is performed by a family of amino acid dehydrogenase (AADH) (Nitta, Yasuda et al. 1974; Oikawa, Yamanaka et al. 2001). The latter catalyzes the reversible transformation between almost all proteinic amino acids and their corresponding oxo acids (Mohammadi, Omidinia et al. 2007),

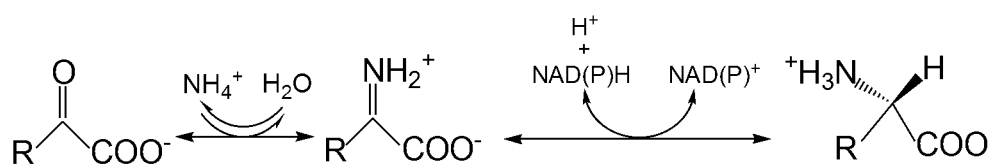

**Scheme 1**

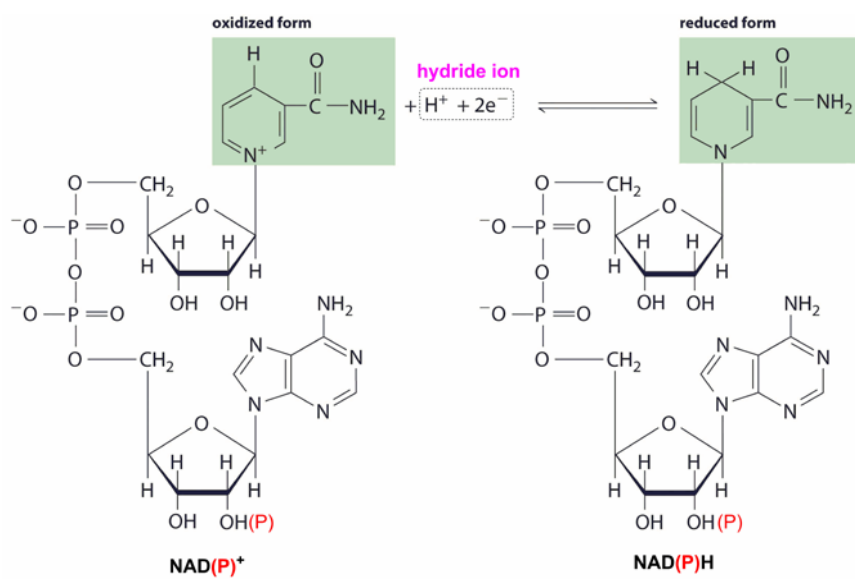

**Scheme 2**

In Scheme 1, the imine intermediate formed between ammonia and α-oxo acids is subsequently reduced by the transfer of a hydride ion (a proton and two electrons) from coenzyme NAD(P)H to form L-amino acids (Scheme 2).

Before this time, it was believed that in the case of enzyme-catalyzed asymmetric synthesis, the specific regiostructure of the active site of enzymes regulates the handedness of the products, just like an auxiliary scaffold or a chiral imprinter(Toone, Werth et al. 1990; Savile, Janey et al. 2010). Here I would suggest an alternative mechanism that the enatioselective synthesis is determined by the spin state of the electron transferred from the co-enzyme NAD(P)H. This mechanism is discussed in detail below.

Spin is an intrinsic property of all fermions whose spin quantum numbers ($m_s$) take half-integer values. For the case of electrons, the spin is divided into two well-entrenched camps, commonly referred to as "spin up" ( ↑ , $m_s = +1/2$) and "spin down" ( ↓ , $m_s = -1/2$). In the absence of a magnetic field, these two states of single electrons or unpaired electrons in atoms are present in statistically equal numbers and degenerate, just like the racemization of two mirror images of chiral molecules. In the presence of magnetic interactions, however, the two spine states can be split and well discriminated (Phipps and Taylor 1927).

Electron spin can also be determined by filling up the electrons in orbital. For example, in the case of reductive amination of α-oxo acids (Eq. 1), we can make product states by describing the spins of two particles, i.e., the π electron of α-C of the imine molecule and the NAD(P)H electron (Fig. 1). Protons and the NAD(P)H electrons react with the Schiff base to produce amino acids. This often happens in the so-called proton-coupled electron transfer (PCET) reactions (Reece and Nocera, 2009). The reduction of the C-H π bond, although needs the transfer of two electrons, proceeds in two successive univalent steps, the intermediate state being a free radical (Guzman Barron 1957; Commoner, Lippincott et al. 1958; Kim, Darley et al. 2008) (Fig. 1). The first electron and a proton (or H atom) will be added to the positive iminium ion, producing a C radical intermediate. After that, there are four possible spin state combination forms of the radical and the second electron (or H atom) since both of them have two spin states (Fig. 1). According to the Pauli exclusion principle, however, two electrons in an orbital can never have same spin. Therefore, pathways (b) and (c) in Fig. 1 is completely forbidden. The rest two pathways (a) and (d) will produce amino acids of different chirality. If channel (a) brought L-amino acids, channel (d) would yield D-amino acids. It should be noted that the spin states of all electrons participating in the reaction are instantaneous; once the chemical bond forms, they will delocalize and their spin orientation will experience transformation due to spin exchange interactions.

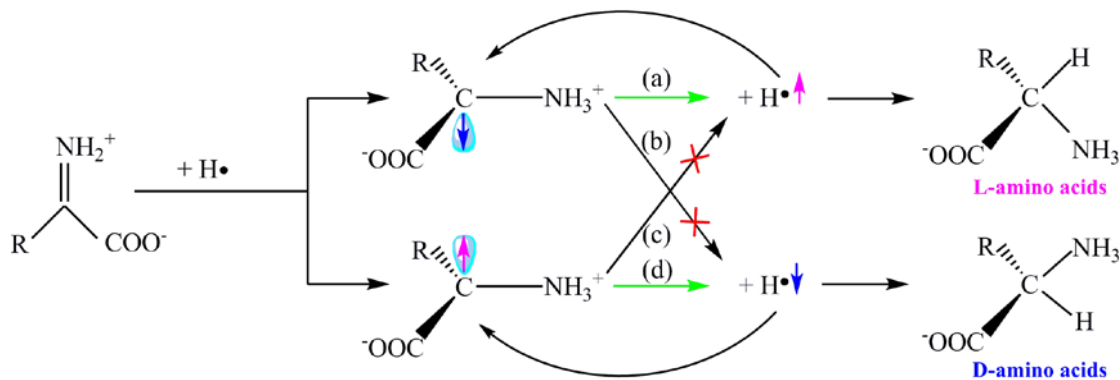

**Fig. 1** Four possible spin state combinations during the formation of amino acids via reductive amination of α-oxo acids.

Since the two imine structures presented in Fig. 1 are totally identical (no chirality), the chirality of the product depends on the spin state of the hydrogen atom. Neutral hydrogen atom consists of a proton and an electron. On account of magnetic interactions between these two particles, a hydrogen atom that has the spins of the electron and proton aligned in the same direction (parallel) has slightly more energy than one where the spins of the electron and proton are in opposite directions (antiparallel) (Rith and Schafer 1999). The two structures are presented in Fig. 2. They are both ground state of hydrogen atom. When they react with the imine molecule, as a result of electron-related magnetic interactions, the antiparallel structure with lower energy may lead to the formation of L-amino acid which has a low energy level than corresponding D-amino acid (Tranter 1987; Macdermott 1995). This energy difference may also help explain why natural enantiomer found in terrestrial biochemistry has intrinsically more stable low energy and prefer over its unnatural counterpart.

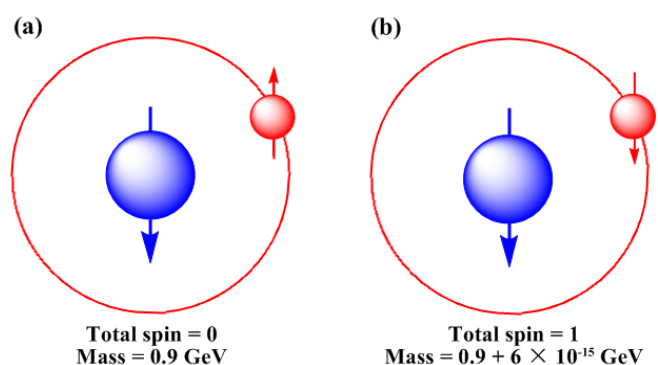

**Fig. 2** In a hydrogen atom, aligning the spins of the proton (blue) and the electron (red) increases the atom's total spin from zero to one. When the electron revolves around the nucleus with antiparallel spin alignment (left), the system occurs in a lower energy state.

According to the principle of minimum energy, all matter in the universe tends to achieve its lowest possible energy state. However, this rule does not apply to the two spin structures of hydrogen atom since they represent two ground states of H atom and the energy gap between them is so small that even very little thermal fluctuation can make them heterogeneous under natural conditions. In this hypothesis paper, therefore, the most critical step in the enzyme-catalytic amino acid synthesis is how a homogenous electron spin state forms.

A better way to find a solution to this problem is keeping a close eye on electron spin property itself. Each spinning electron causes a magnetic field to form around it. Electrons with different spin orientations generate magnetic fields in opposite directions. So an external magnetic field can make electron spin state split through magnetic interactions. However, only a few solid materials which own very large spin-orbit coupling (SOC) are naturally magnetic. In pure organic compounds, electrons in all atoms exist in pairs, with each electron spinning in an opposite direction. The magnetic field of one electron is cancelled by an opposite magnetic field produced by the other electron in the pair. Thus SOC in organic matter is commonly believed to be too weak to generate a magnetic field. However, theoretical and experimental studies demonstrated that close-packed well-organized chiral molecular layers do have appreciable magnetic properties (Carmeli, Skakalova et al. 2002; Vager and Naaman 2002). The mechanism of these phenomena is not clear yet. But it may deal with the molecular chirality.

Proteins and enzymes are such chiral biopolymers because of the asymmetry of their monomeric building blocks (i.e., L-amino acids) and helical secondary structure (mostly right-handed). Amino acid monomers are dipolar molecules. In proteins, when they are assembled together via peptide bonds, the original dipole directions of each monomer are packed parallel in one direction. Therefore, particular arrangements of biomolecules such as α-helix have large macro-dipoles (Fig. 3), which induce strong electric fields (Wada 1976; Hol, van Duijnen et al. 1978; Hol 1985). On the other hand, in amino acid molecules, each covalent bond has stretching vibration. When $C_\alpha$-H bond contracts or stretches, nuclear motion can induce charge redistribution, producing charge flows along bonds. This will result in the formation of a nonzero electric dipole transition moment and a nonzero magnetic dipole transition moment in amino acid molecules (Nafie, Oboodi et al. 1983). The amide group has a partial double-bond character, which gives the peptide unit two resonance forms (Fig .4a). With resonance, the nitrogen is able to donate its lone pair of electrons to the carbonyl carbon and push electrons from the carbonyl double bond towards the oxygen, forming the oxygen anion which is estimated at 40% under typical conditions. This resonance effect is very stabilizing because the electrons can be delocalized over multiple atoms. From Fig. 4b and Fig. 4c, it can be found that whether the $C_\alpha$-H bond contracts or stretches, electrons move from the carboxyl terminus (C-terminus) to the amino terminus (N-terminus), just the same as in the dipolar electric field in Fig. 3.

Therefore, when a charge is captured by the electric field and moves within the macromolecular system in one direction, it creates a magnetic field, just like a classical electromagnetic coil. Owing to the broken mirror image symmetry of amino acids, the magnetic field also occurs only in one direction (Fig. 5). In such a magnetic field with broken symmetry, the spine states of an electron can be split and well discriminated.

This is not a groundless assumption. Recently, a series of excellent experimental work by Professor Naaman's group (Ray, Ananthavel et al. 1999; Ray, Daube et al. 2006; Gohler, Hamelbeck et al. 2011; Naaman and Vager 2011; Xie, Markus et al. 2011; Naaman and Waldeck 2012) has pioneeringly demonstrated that chiral biopolymers such as double strand DNA and α-helical peptide can act as an organic spin filter to produce spin polarized electron. This phenomenon is called chiral-induced spin selectivity (CISS).

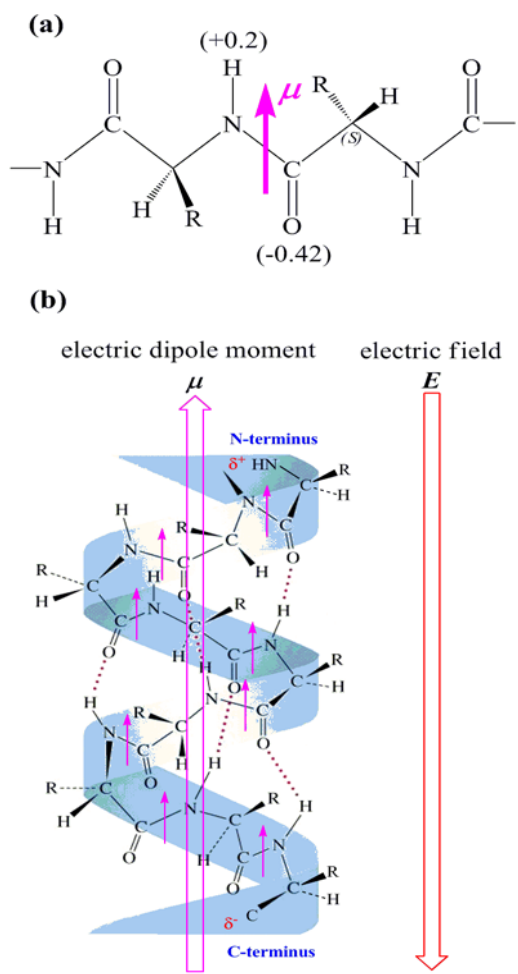

**Fig. 3** Geometry and dipole moments of the peptide unit (a) and the α-helix structure (b).

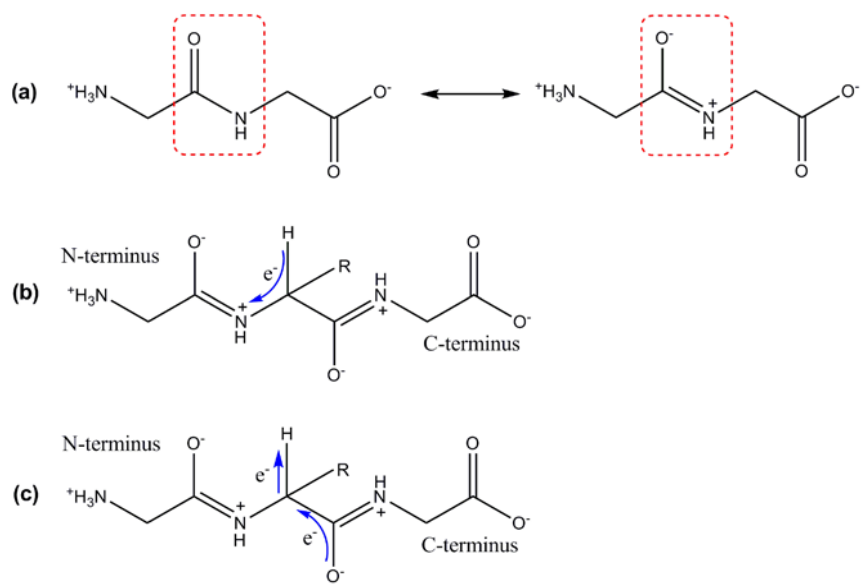

**Fig. 4** (a) Two resonance forms of the amido bond in peptides; (b) and (c) $C_\alpha$-H bond stretching vibration induce charge redistribution in the peptide chain.

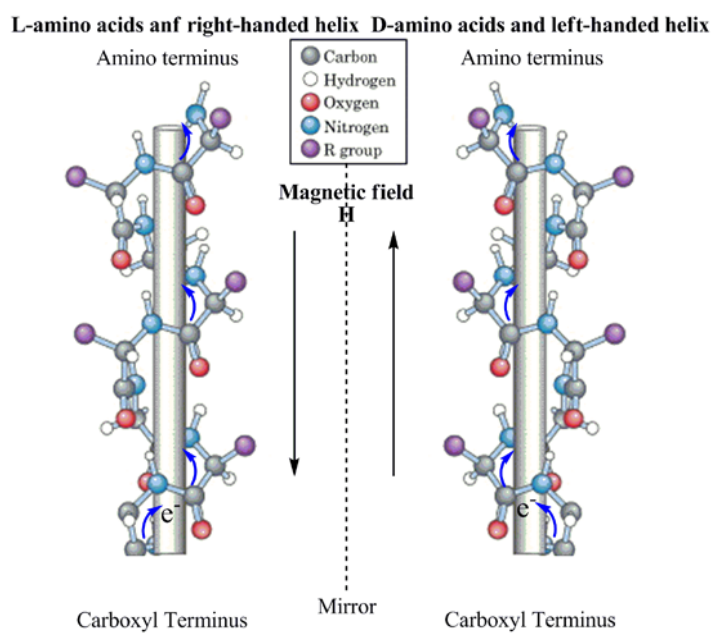

**Fig. 5** The electron motions in L-amino acids constructed right-handed α-helix and D-amino acids constructed left-handed α-helix.

Thus far, from the deliberation above, the hypothesis that I want to present in this paper is getting clearer and clearer. I would expect that the emergence of a single chiral form of biomolecules in living organisms is specifically determined by the electron spin state during their enzyme-catalyzed synthesis. A phenomenological model is represented in Fig. 6 and described as follows. The electrons released from NAD(P)H are heterogeneous in spin states; however, when passing through the chiral α-helix structure to the site of amino acid synthesis at the other end of the helix, they experience the electrostatic potential of the molecule, and also the inherent magnetic field or the magnetic field produced by themselves; their spin states are filtered and polarized, producing only "spin up" electrons; once the spin-polarized electrons participate the reductive reaction of the imine molecules, only L-amino acids are formed according to the Pauli exclusion principle (Fig. 1).

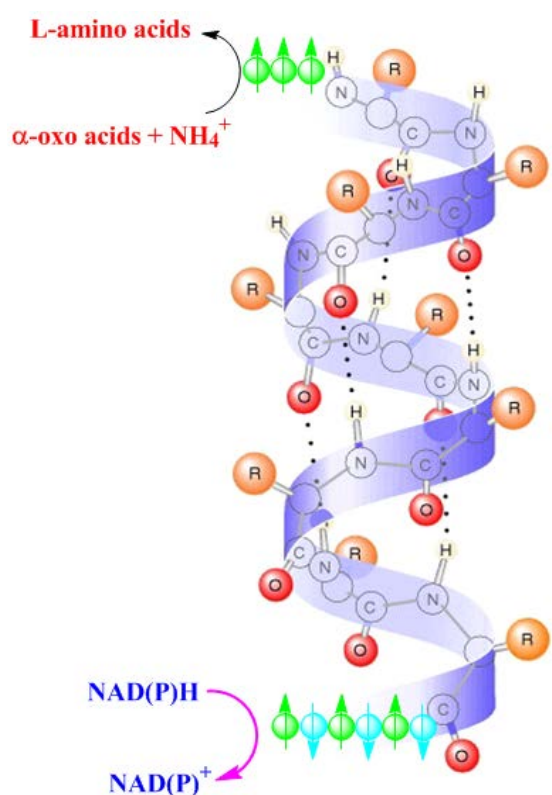

**Fig.6** A phenomenological model for the enzyme-catalyzed chiral amino acid synthesis.

## Postscript

In this paper, I tentatively put forward a mental picture for the enzyme-assisted asymmetric amino acid synthesis in living bodies. Then how could a homochirality world of biomolecules have formed in the absence of enzymatic networks before the origins of life? This will be discussed in a paper in preparation entitled "Electron spin and the origin of bio-homochirality II. Prebiotic inorganic reaction model". In that paper, I will talk about the electron spin properties in transitional metal sulfide, such as ferromagnetic $Fe_3S_4$ and doped ZnS diluted magnetic semiconductor, and their possible role in the prebiotic synthesis of L-amino acids, based on the hydrothermal vent theory for the origins of life (Wang, Yang et al. 2011; Wang, Li et al. 2012).